\pdfoutput=1

\documentclass{nature}

\usepackage{url}
\usepackage{graphicx}

\makeatletter
\let\saved@includegraphics\includegraphics
\AtBeginDocument{\let\includegraphics\saved@includegraphics}
\renewenvironment*{figure}{\@float{figure}}{\end@float}
\makeatother

\spacing{1}

\title{Direct measurement of stellar angular diameters by the VERITAS Cherenkov Telescopes}

\author{
W.~Benbow$^{1}$,
R.~Bird$^{2}$,
A.~Brill$^{3}$,
R.~Brose$^{4,5}$,
A.~J.~Chromey$^{6}$,
M.~K.~Daniel$^{1,*}$,
Q.~Feng$^{3}$,
J.~P.~Finley$^{7}$,
L.~Fortson$^{8}$,
A.~Furniss$^{9}$,
G.~H.~Gillanders$^{10}$,
C.~Giuri$^{5}$, 
O.~Gueta$^{5}$, 
D.~Hanna$^{11}$,
J.~Halpern$^{3}$, 
T.~Hassan$^{5,*}$,
J.~Holder$^{12}$,
G.~Hughes$^{1}$,
T.~B.~Humensky$^{3}$, 
A.~M.~Joyce$^{10}$,
P.~Kaaret$^{13}$,
P.~Kar$^{14}$,
N.~Kelley-Hoskins$^{5}$,
M.~Kertzman$^{15}$,
D.~Kieda$^{14}$,
M.~Krause$^{5}$,
M.~J.~Lang$^{10}$,
T.~T.~Y.~Lin$^{11}$,
G.~Maier$^{5}$,
N.~Matthews$^{14}$, 
P.~Moriarty$^{10}$,
R.~Mukherjee$^{16}$, 
D.~Nieto$^{3, 23}$, 
M.~Nievas-Rosillo$^{5}$, 
S.~O'Brien$^{17}$,
R.~A.~Ong$^{2}$,
N.~Park$^{18}$,
A.~Petrashyk$^{3}$,
M.~Pohl$^{4,5}$,
E.~Pueschel$^{5}$,
J.~Quinn$^{17}$,
K.~Ragan$^{11}$,
P.~T.~Reynolds$^{19}$,
G.~T.~Richards$^{12}$,
E.~Roache$^{1}$, 
C.~Rulten$^{8, 24}$, 
I.~Sadeh$^{5}$,
M.~Santander$^{20}$,
G.~H.~Sembroski$^{7}$,
K.~Shahinyan$^{8}$, 
I.~Sushch$^{4}$, 
S.~P.~Wakely$^{21}$,
R.~M.~Wells$^{6}$,
P.~Wilcox$^{13}$,
A.~Wilhelm$^{4,5}$,
D.~A.~Williams$^{22}$,
T.~J.~Williamson$^{12}$
}

\begin{document}

\maketitle

\begin{affiliations}
\item{Center for Astrophysics $|$ Harvard \& Smithsonian, Fred Lawrence Whipple Observatory, Amado, AZ 85645, USA} 
\item{Department of Physics and Astronomy, University of California, Los Angeles, CA 90095, USA} 
\item{Physics Department, Columbia University, New York, NY 10027, USA} 
\item{Institute of Physics and Astronomy, University of Potsdam, 14476 Potsdam-Golm, Germany} 
\item{DESY, Platanenallee 6, 15738 Zeuthen, Germany} 
\item{Department of Physics and Astronomy, Iowa State University, Ames, IA 50011, USA} 
\item{Department of Physics and Astronomy, Purdue University, West Lafayette, IN 47907, USA} 
\item{School of Physics and Astronomy, University of Minnesota, Minneapolis, MN 55455, USA} 
\item{Department of Physics, California State University - East Bay, Hayward, CA 94542, USA} 
\item{School of Physics, National University of Ireland Galway, University Road, Galway, Ireland} 
\item{Physics Department, McGill University, Montreal, QC H3A 2T8, Canada} 
\item{Department of Physics and Astronomy and the Bartol Research Institute, University of Delaware, Newark, DE 19716, USA} 
\item{Department of Physics and Astronomy, University of Iowa, Van Allen Hall, Iowa City, IA 52242, USA} 
\item{Department of Physics and Astronomy, University of Utah, Salt Lake City, UT 84112, USA} 
\item{Department of Physics and Astronomy, DePauw University, Greencastle, IN 46135-0037, USA} 
\item{Department of Physics and Astronomy, Barnard College, Columbia University, NY 10027, USA} 
\item{School of Physics, University College Dublin, Belfield, Dublin 4, Ireland} 
\item{WIPAC and Department of Physics, University of Wisconsin-Madison, Madison WI, USA} 
\item{Department of Physical Sciences, Cork Institute of Technology, Bishopstown, Cork, Ireland} 
\item{Department of Physics and Astronomy, University of Alabama, Tuscaloosa, AL 35487, USA} 
\item{Enrico Fermi Institute, University of Chicago, Chicago, IL 60637, USA} 
\item{Santa Cruz Institute for Particle Physics and Department of Physics, University of California, Santa Cruz, CA 95064, USA} 
\item{Now at Universidad Complutense de Madrid, Facultad de Ciencias Físicas, Plaza Ciencias 1, E-28040 Madrid, Spain}
\item{Now at Department of Physics, University of Durham, Durham, DH1 3LE, UK.}
\end{affiliations}

\begin{abstract}
The angular size of a star is a critical factor in determining its basic properties\cite{Mozurkewich2003}.
Direct measurement of stellar angular diameters is difficult: at interstellar distances stars are generally too small to resolve by any individual imaging telescope.
This fundamental limitation can be overcome by studying the diffraction pattern in the shadow cast when an asteroid occults a star\cite{Roques87}, but only 
when the photometric uncertainty is smaller than the noise added by atmospheric scintillation\cite{Morbey74}.
Atmospheric Cherenkov telescopes used for particle astrophysics observations have not generally been exploited for optical astronomy due to the modest optical quality of the mirror surface.
However, their large mirror area makes them well suited for such high-time-resolution precision photometry measurements\cite{Lacki2011}. 
Here we report two occultations of stars observed by the VERITAS\cite{VERITAS} Cherenkov telescopes with millisecond sampling, from which we are able to provide a direct measurement of the occulted stars' angular diameter at the $\leq0.1$\,milliarcsecond scale.
This is a resolution never achieved before with optical measurements and represents an order of magnitude improvement over the equivalent lunar occultation method\cite{Ridgway77}.
We compare the resulting stellar radius with empirically derived estimates from temperature and brightness measurements, confirming the latter can be biased for stars with ambiguous stellar classifications.

\end{abstract}

When a solar-system object, such as an asteroid or the Moon, passes in front of a star as viewed on the celestial sphere, it provides a powerful tool for studying both the occulting object and the occulted star\cite{Roques87}. 
As viewed from the ground, the rapid drop in the observed intensity of light is modified by diffraction fringes preceding/following the edges of the central shadow region of the obscuring object. 
Above a minimum angular size\cite{Ridgway77}, the extended disc of a star will modify and reduce the intensity of the diffraction fringes, diverging noticeably from the pattern of a point-like source, until it reaches angular diameters where the background object is fully geometrically resolved and the diffraction fringes disappear. 
A fit to observable diffraction fringes thereby enables a direct measurement of the angular size of the star, even though this may be far below the imaging angular resolution limit of the telescope.
Observations of stellar occultations by asteroids are frequently used to determine the properties of an asteroid\cite{iota_web} (size, shape) and are also theoretically capable of angular size measurements well below the 1\,milliarcsecond (mas) scale that has ultimately been a limit to the lunar occultation technique. 
In fact, benefitting from the increased distance to the occulting surface with respect to the Moon, they should have an even smaller potentially achievable resolution. 
However, to date, there has been little success in measuring asteroid occultation diffraction fringes to make such angular size measurements.

\begin{figure}[th!]
\begin{center}
\includegraphics[width=0.405\linewidth]{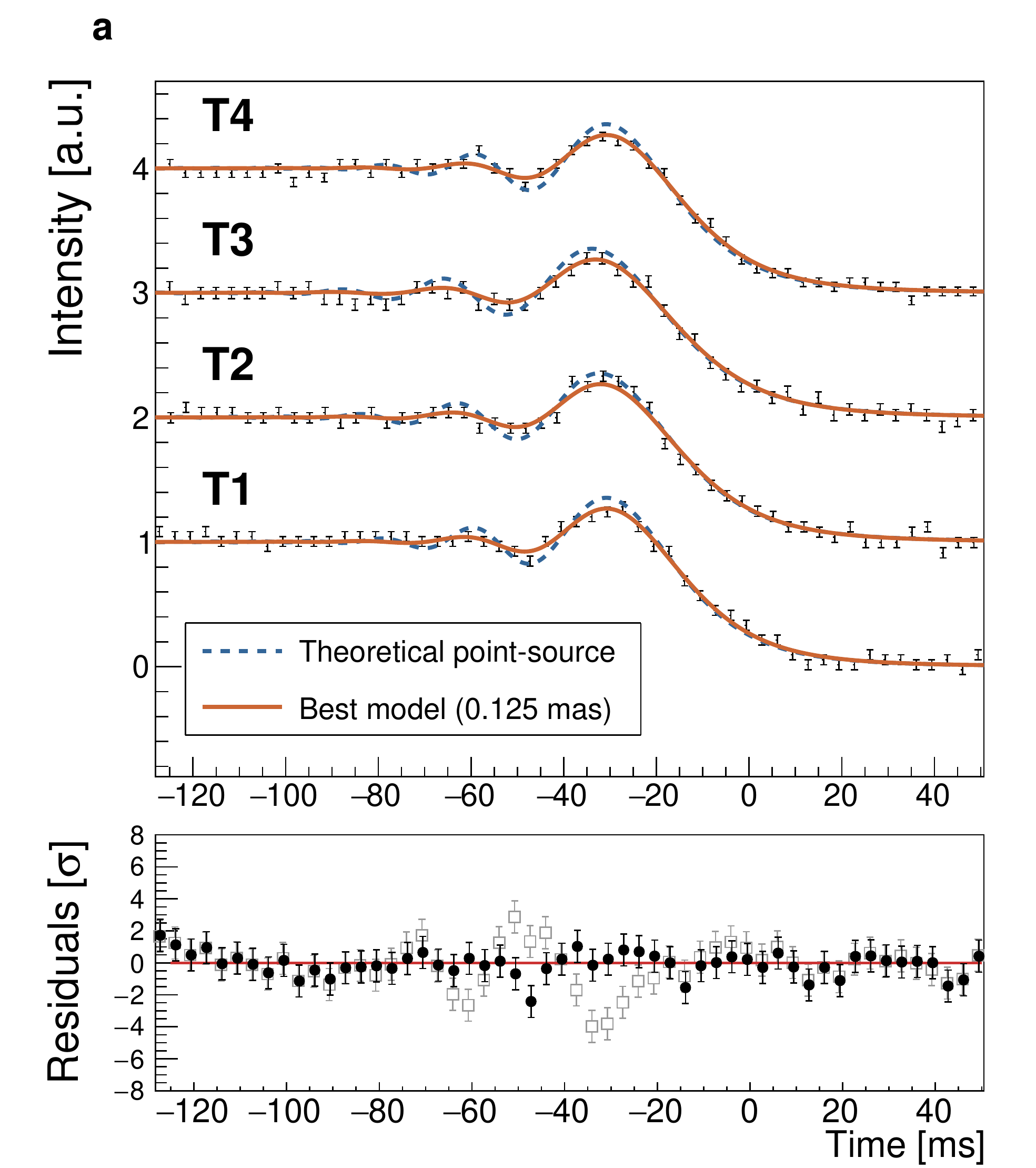}
\includegraphics[width=0.405\linewidth]{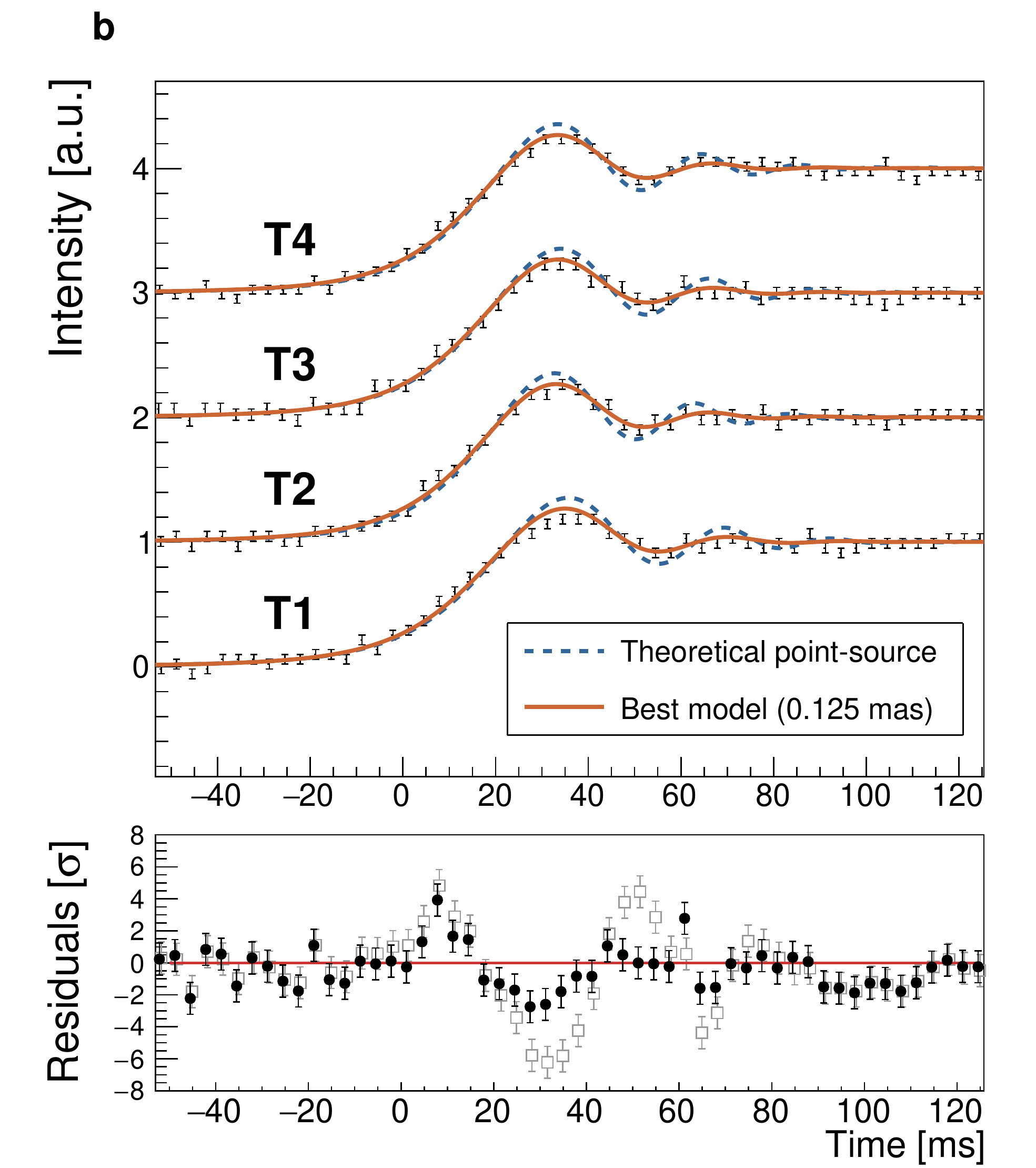}\\
\includegraphics[width=0.405\linewidth]{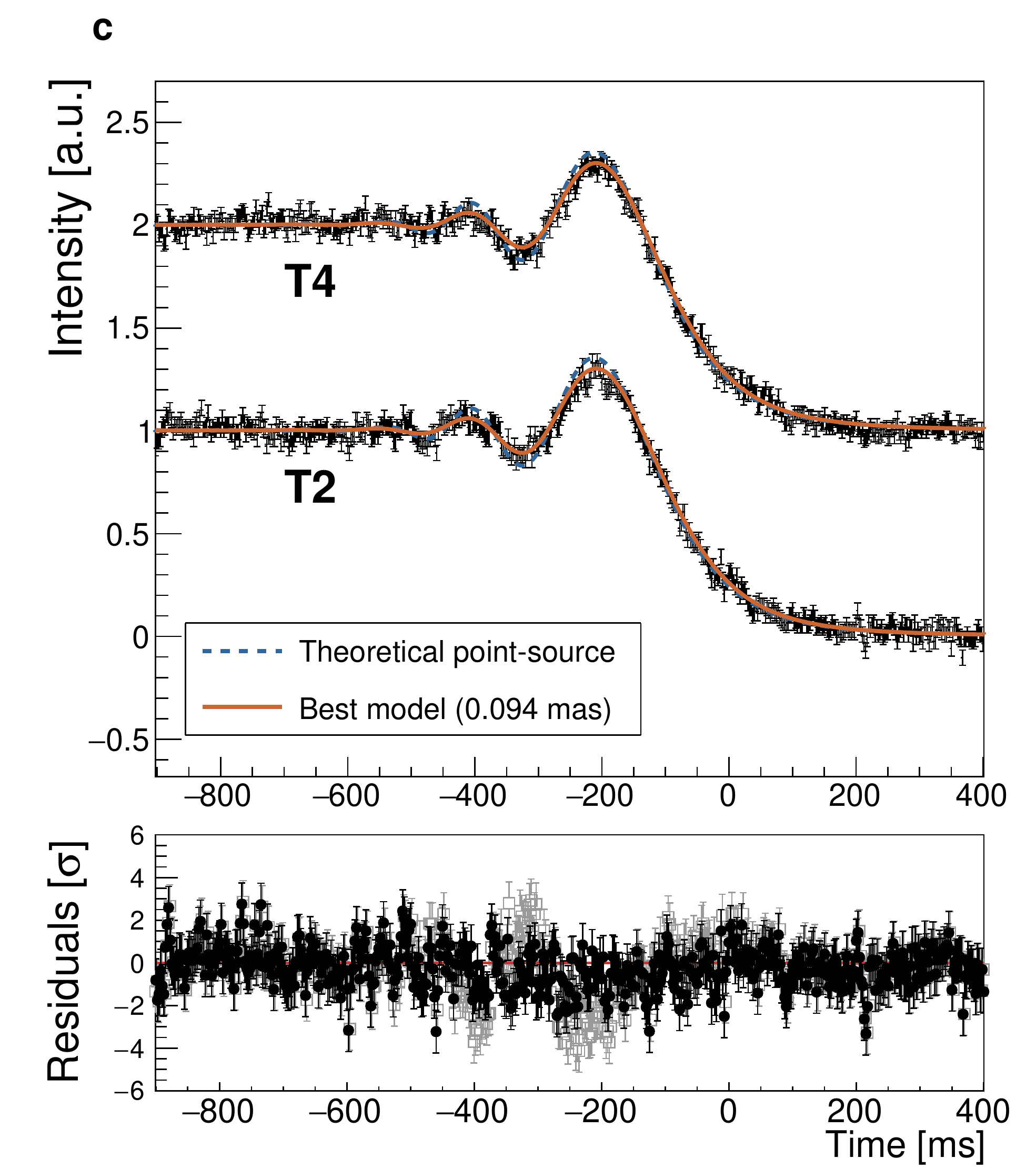}
\includegraphics[width=0.405\linewidth]{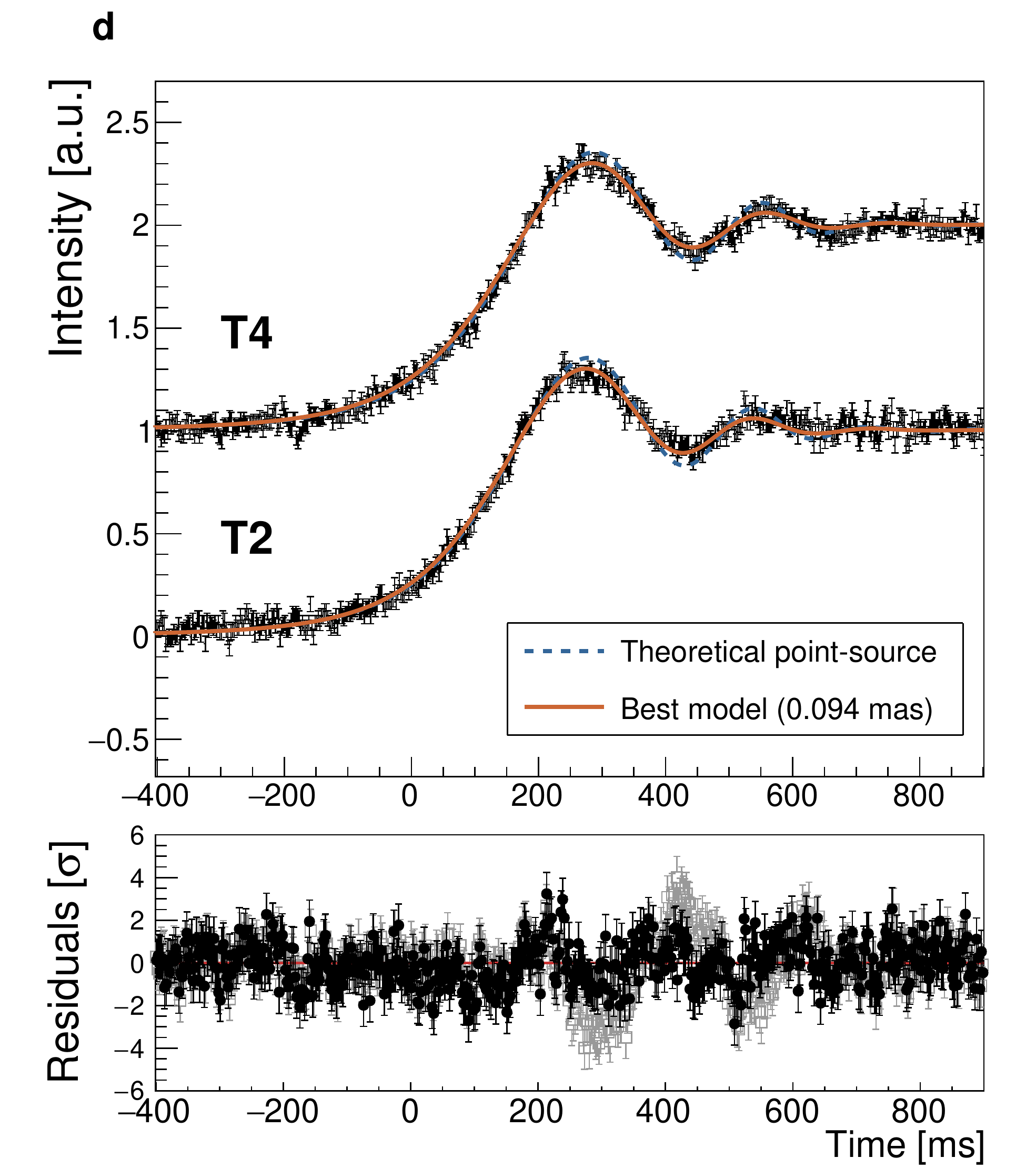}
\end{center}
\caption{\textbf{Ingress and egress light curves for both asteroid occultations.} \textbf{a}: The light curves of the ingress of the (1165) Imprinetta / TYC\,5517-227-1 occultation, with the best-fit diffraction pattern (red line) and theoretical point-source model (dashed blue line). 
Each telescope light curve is normalized such that the unocculted and occulted intensity levels correspond to 1 and 0 respectively, with an added y-axis offset between telescopes for clarity.
The combined (averaged) residual with respect to the point-source (grey empty squares) and best-fit (black filled circles) models are shown in the bottom panels. Vertical error bars are defined as the 68\% containment radius, including systematics.
\textbf{b}: The same for the egress of the (1165) Imprinetta / TYC\,5517-227-1 occultation. 
\textbf{c}: The same for the ingress of the (201) Penelope / TYC\,278-748-1 occultation.
\textbf{d}: The same for the egress of the (201) Penelope / TYC\,278-748-1 occultation.
}
\label{fig:Shadow}
\end{figure}

On 22nd February, 2018 the 
asteroid (1165)\,Imprinetta occulted the 10.2 V-magnitude (m$_V$) star TYC\,5517-227-1, with the shadow path predicted to have a 50\% chance of detection from the Fred Lawrence Whipple Observatory (FLWO), where the Very Energetic Radiation Imaging Telescope Array System (VERITAS) is sited. The four 12\,m diameter imaging atmospheric Cherenkov telescopes (IACTs) of VERITAS act as effective ``light buckets'' to collect the fast, faint emission of Cherenkov light generated by particle air showers initiated in the upper atmosphere by very-high-energy cosmic radiation. 
This large mirror surface also makes VERITAS a very sensitive detector for  high-time-resolution optical photometry following a recent upgrade of the camera's central pixel monitoring instrumentation (see Methods for details).
Distinct diffraction fringes were detected during ingress and egress, as shown in Figure~\ref{fig:Shadow}~\textbf{a},\textbf{b} respectively. 
This marks the first time an occultation has been measured using an IACT and successfully demonstrates that these instruments are indeed capable photometers for optical astronomy. 

\begin{figure}[bh!]
\begin{center}
\includegraphics[width=0.45\linewidth]{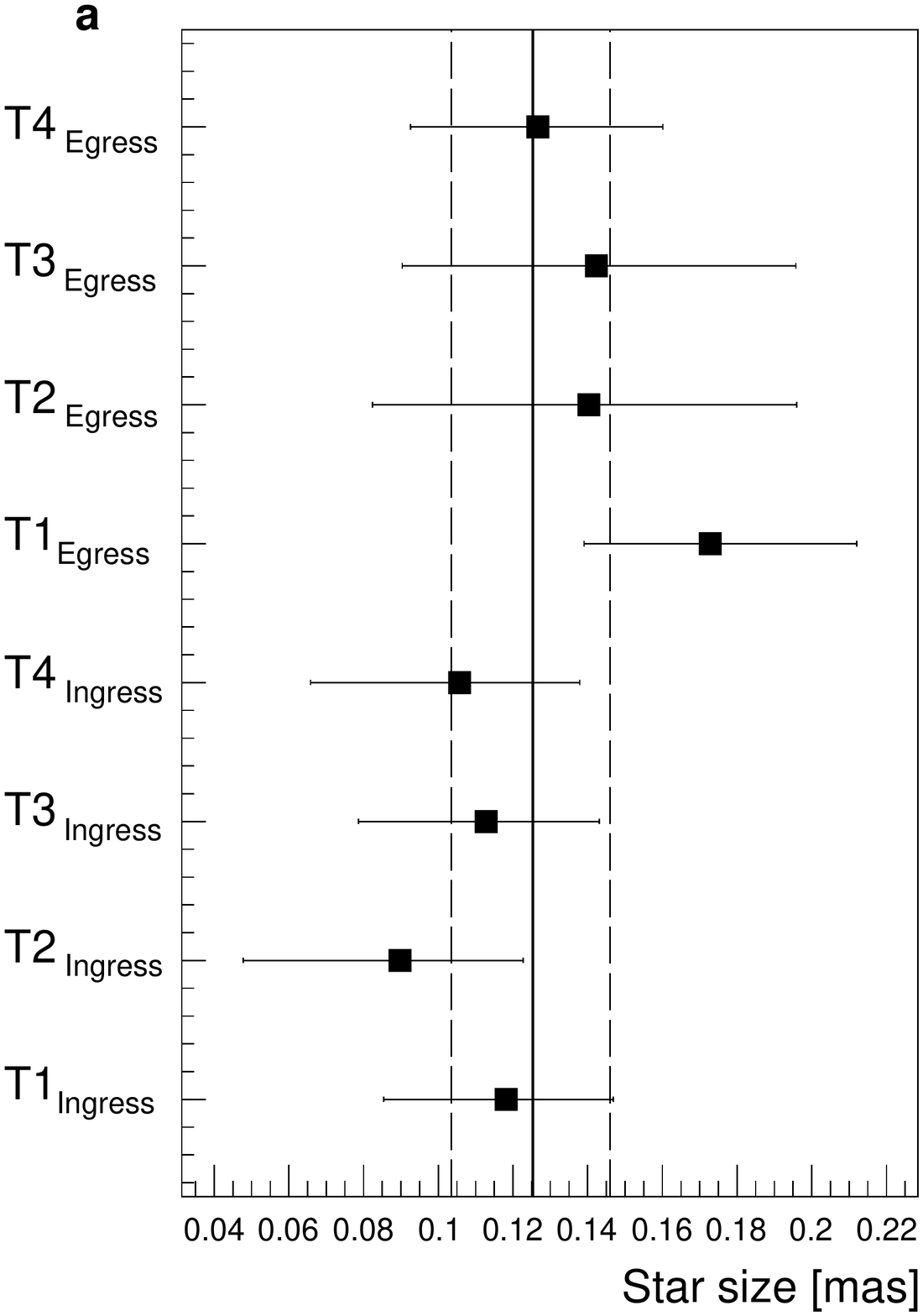} 
\includegraphics[width=0.45\linewidth]{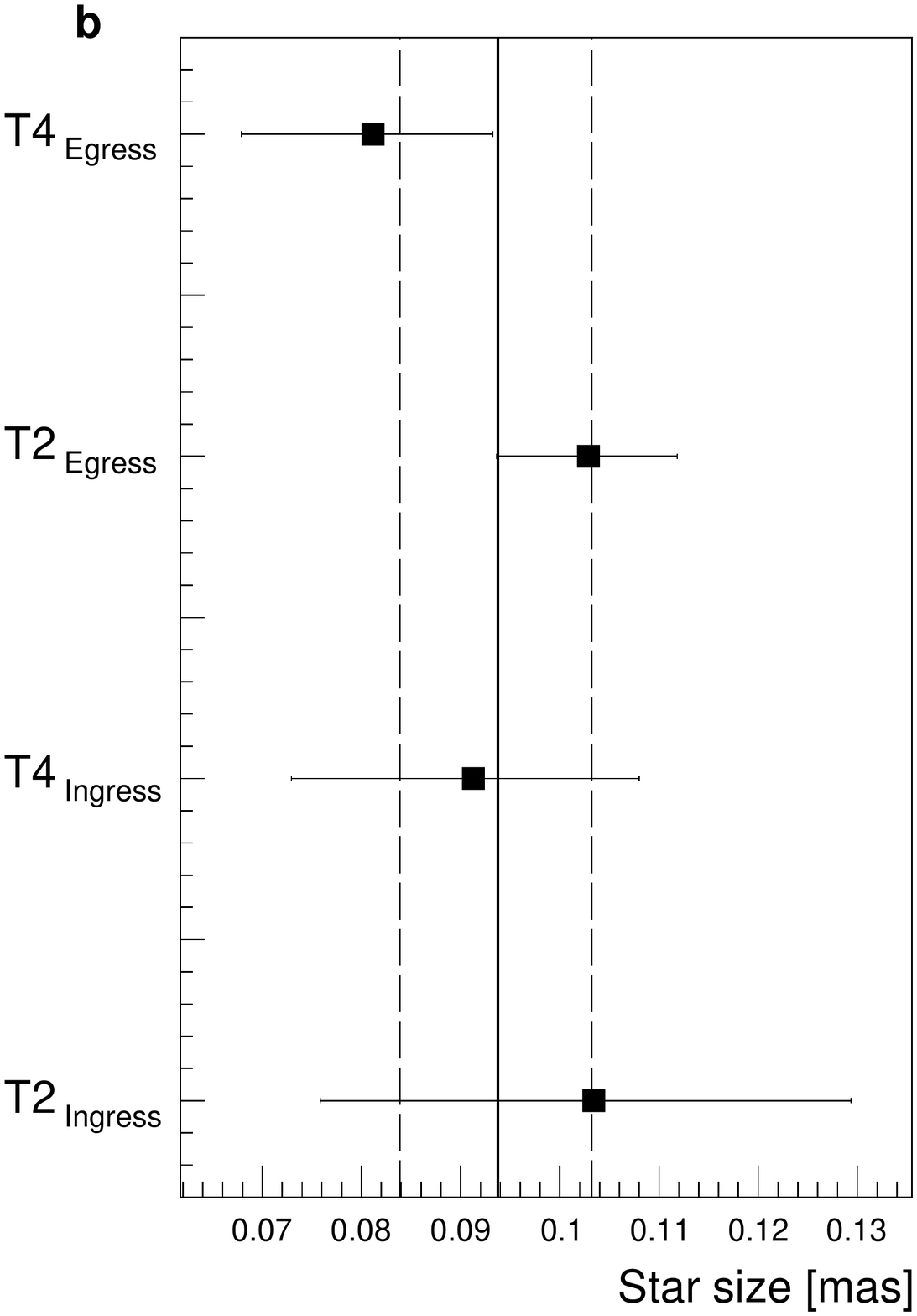} 
\end{center}
\caption{\textbf{Individual and combined stellar size measurements.}
\textbf{a}: Stellar size measurements of TYC\,5517-227-1 from each ingress and egress light curve assuming an uniform disc profile. 
Vertical lines show the final stellar size measurements with their 1-$\sigma$ uncertainty regions (dashed lines), obtained from the combined $\chi^2$ minimisation of all available measurements (see Methods section for details). 
Horizontal error bars refer to the 68\% containment radius associated to each measurement individually. 
\textbf{b}: the same for TYC\,278-748-1. 
Only two of the telescopes were used for this analysis. 
}
\label{fig:Fit}
\end{figure}

\begin{figure}[hb]
\includegraphics[width=0.51\linewidth]{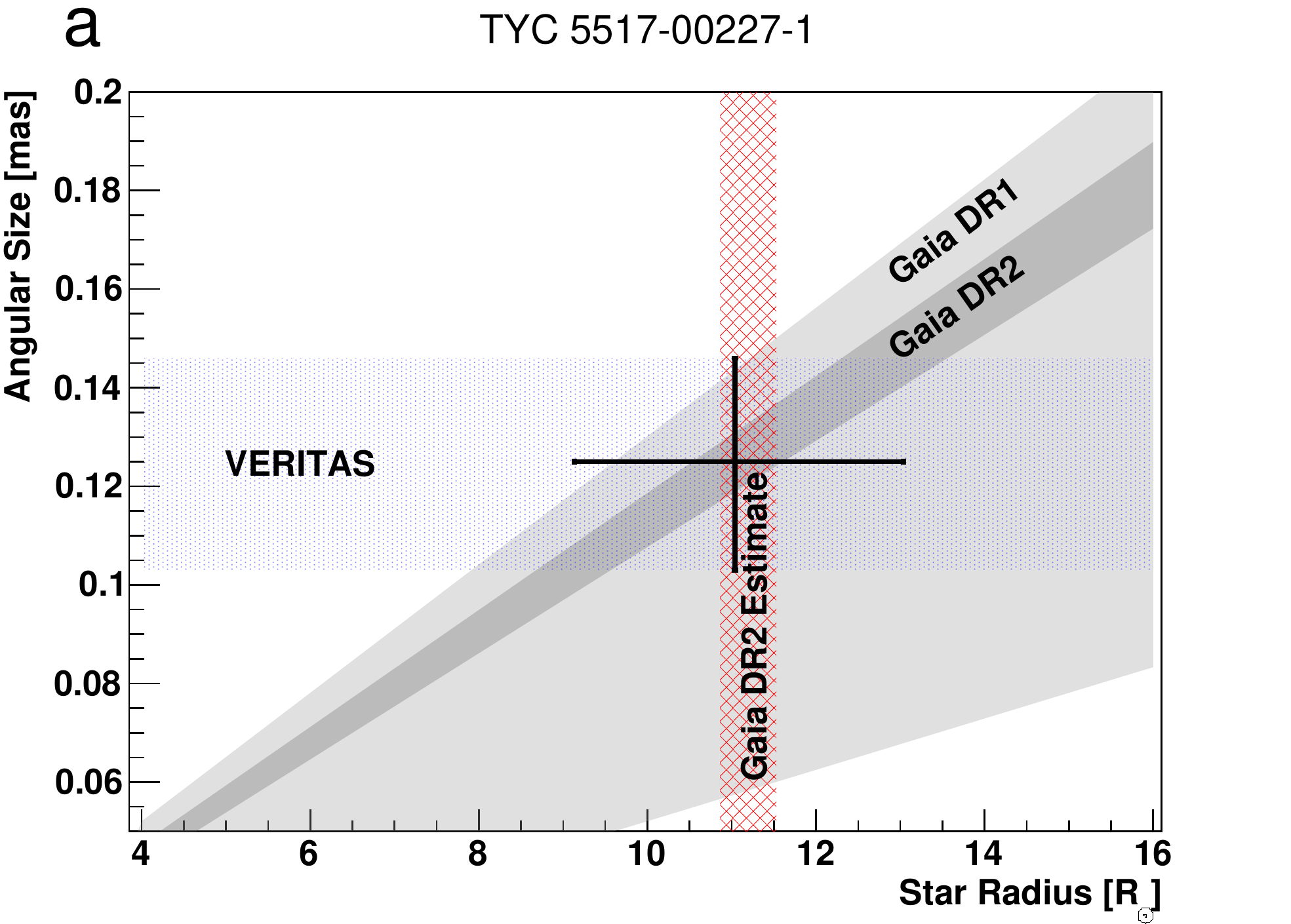}
\includegraphics[width=0.51\linewidth]{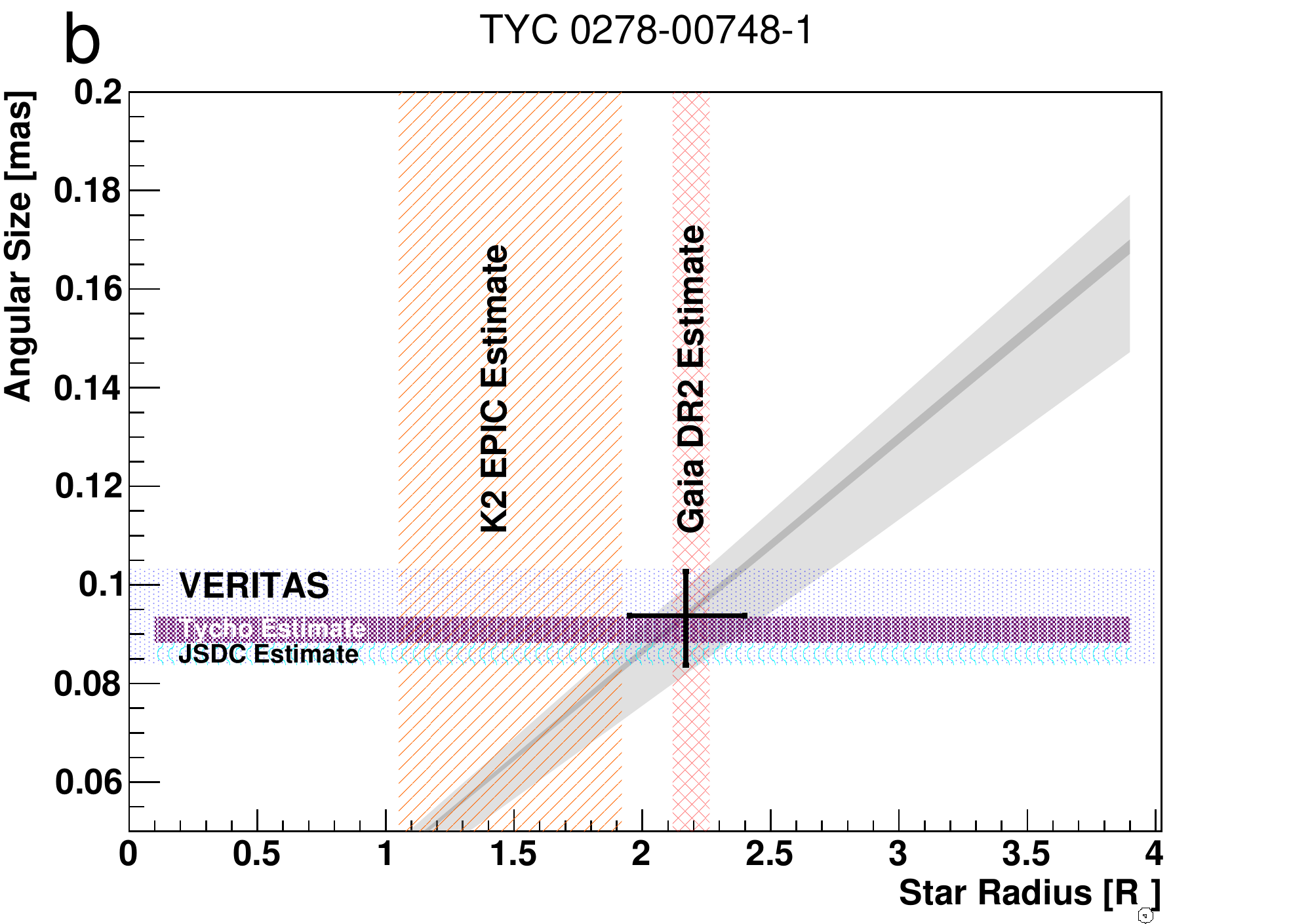}
\caption{\textbf{Comparison of the angular size measurements and stellar radius estimates in this work compared to those available in the literature.} 
\textbf{a}: Angular size as a function of stellar radius for the parallax distance of TYC\,5517-227-1 as determined by Gaia Data Release 1 (light grey band) and Gaia Data Release 2 (dark grey band). 
A model estimate of the stellar radius from Gaia~FLAME is shown by the red hatched box, the best-fit angular size from our measurements is shown by the blue stippled box, and the estimated uniform disc approximation stellar radius for our best fit angular size at the Gaia DR2 parallax distance is marked by a cross.
\textbf{b}: The same for TYC\,278-748-1, with the addition of the angular size estimates from the Tycho\cite{Tycho} (magenta box) and JSDC\cite{JSDC} (cyan box) catalogues and the radius estimate from the Kepler~K2~EPIC\cite{K2EPIC} catalogue (orange diagonal-stripe box).
}
\label{fig:Size}
\end{figure}

Knowing the distance and velocity of the asteroid, and accounting for the optical bandpass of the detected photons, allows us to find the stellar angular size that best fits the observation, assuming a given radial intensity profile of the occulted star. 
The interference of different wavelengths of light accepted by the detector also reduces the intensity of the diffraction fringes, representing the largest systematic uncertainty to the size estimate. 
However, the high signal-to-noise ratio provided by the large light-collection area of the IACT mirrors and the multiple independent measurements provided by each telescope allow us to discriminate with high confidence the effect of the star size even with a wide optical bandpass ($\sim120$\,nm) photodetector\cite{Ridgway77,TDI}. 
At the time of the occultation the exact classification of the star remained somewhat ambiguous from spectral data alone. 
Follow-up observations with the Michigan-Dartmouth-MIT (MDM) Observatory determined its spectral type\cite{hammer} to be K3, either main sequence or an evolved giant. 
The diffraction pattern measured by VERITAS constrains the uniform disc approximation of the star's angular size to be $0.125^{+0.021}_{-0.022}$\,mas, as shown in Figure~\ref{fig:Fit}\textbf{a}. 
Once the measured parallax distance\cite{Gaia2018b} of $820\pm40$\,parsecs (pc) is taken into account, the angular size measurement determines the radius to be $11.0^{+1.9}_{-2.0}\,\mathrm{R_{\odot}}$, as shown in Figure~\ref{fig:Size}\textbf{a}, which when combined with the effective temperature from the spectral measurement unambiguously establishes it to be a K3III giant star. 
Limb or gravity darkening\cite{Claret2011,Howarth2011} would cause the true radius of the star to be slightly larger than the uniform disc value, but by an amount that is smaller than the bounds of the measurement uncertainty we obtain here (typically below the 10\% level\cite{NeilsonLesterI,NeilsonLesterII}). 

In Figure~\ref{fig:Fit}\textbf{a} there is some, but not a significant, hint of a mismatch between the best fitting model of ingress and egress measurements, possibly indicating both sides of the asteroid may not be equally well modelled by an identical straight-edge assumption, and thus potentially implying a potential systematic difference between the two edges. 
The effect of an irregular occulting surface has been studied in the context of lunar occultations\cite{Morbey74, LimbIrregularities} with the conclusion that such effects could be significant in special configurations, albeit generally unlikely. 
If we assume that any surface irregularities on an asteroid would act to distort the diffraction pattern in a similar fashion to those on the lunar limb, then at a sufficient scale they will also tend to modify the fringe intensity\cite{LimbIrregularities}, which in turn leads to a mis-estimation of the stellar angular size.
Taking our optical bandpass and assuming the features to be perpendicular to the line of sight means we can constrain any surface irregularities at the level of $\leq$3\% of the asteroid radius.
Although it is possible, indeed probable, that irregular limb features at this level would be present on (1165) Imprinetta, it is unlikely that we would be able to retrieve the exact limb profile from these data alone as fit solutions are unlikely to be unique\cite{LimbIrregularities}. 
Any mismatch between the data and considered diffraction models is unlikely to be due to different components of a stellar binary system being resolved as there is no corresponding step function in the lightcurve that would be the characteristic signature of multiple components in a system. 

Following the success of the Imprinetta observation, on 22nd May, 2018 an occultation of a 9.9 m$_V$ star, TYC\,278-748-1, by the 88\,km diameter asteroid (201)\,Penelope was observed with a predicted 29\% chance the shadow zone could pass over FLWO. 
Again, the diffraction pattern was clearly detected (see Figure~\ref{fig:Shadow}\textbf{c},\textbf{d}) and the star's angular size directly measured to be $0.094^{+0.009}_{-0.010}$\,mas, as shown in Figure~\ref{fig:Fit}\textbf{b}. 
This is consistent, within errors, to uniform disc angular size estimates from the Tycho\cite{Tycho} and JSDC\cite{JSDC} catalogues. With the measured parallax distance\cite{Gaia2018b} of $215\pm2$\,pc, we establish the star to have a directly determined radius of $2.17^{+0.22}_{-0.23}\,\mathrm{R_{\odot}}$. 
The measured effective temperature\cite{Gaia2018b} of $5768^{+74}_{-115}$\,K, places the spectral classification as G (similar in spectrum to the Sun, a G2V).
The only available estimates of the radius to make such a classification are from empirical fits to measurements of the effective temperature and  luminosity in the Kepler~K2 Ecliptic Plane Input Catalog (EPIC)\cite{K2EPIC} and Gaia~DR2 Final Luminosity, Age and Mass Estimator (FLAME)\cite{Gaia2018b} catalogue, at $1.415^{+0.503}_{-0.371}\,\mathrm{R_\odot}$ and 
$2.173^{+0.055}_{-0.089}\,\mathrm{R_\odot}$ respectively. 
The radius we measure places it as a sub-giant (IV), clearly favouring the Gaia~DR2 estimates in a manner that is independent of the degeneracies in the assumptions associated with inferred radius estimates\cite{Andrae2018}. 
As the K2~EPIC targets are known to mis-classify a significant fraction of subgiants as dwarfs\cite{Tycho}, hence systematically underestimating their radii, this is entirely consistent with our findings.
This knowledge does impact the choice of database to use, for instance, in estimating the size of transiting exoplanets from the radius of the host star\cite{Boyajian2015}, with Gaia~DR2 appearing more reliable than the Kepler estimates.

The diffraction pattern fitting technique has been successfully exploited with lunar occultation measurements\cite{Ridgway77, TDI, JMDC} to measure stellar angular diameters down to the $\sim$1\,mas level, particularly toward the red end of the optical spectrum ($\lambda>600$\,nm) where background light from the Moon is minimised. 
Up to now, measurements of stellar angular diameters below 1\,mas have instead been largely reliant on interferometric measurements. 
Amplitude interferometry\cite{KStarsCHARA,NPOI,JMDC} observations are again largely limited to the redder end of the spectrum due to atmospheric scintillation noise effects ultimately limiting the ability to correct the optical path length to the necessary fraction of a wavelength. 
Intensity interferometry\cite{SII,JMDC} is an alternative method free from scintillation noise and so able to extend into the blue end of the optical spectrum, but very large mirror surfaces are also required and the technique is intrinsically limited to only the measurement of bright, hot sources (historically m$\leq3$, T$\geq 10,000$\,K, but more sensitive instruments are in development\cite{CTAWP}). 

\begin{figure}[!ht]
\begin{center}
\includegraphics[width=0.66\linewidth]{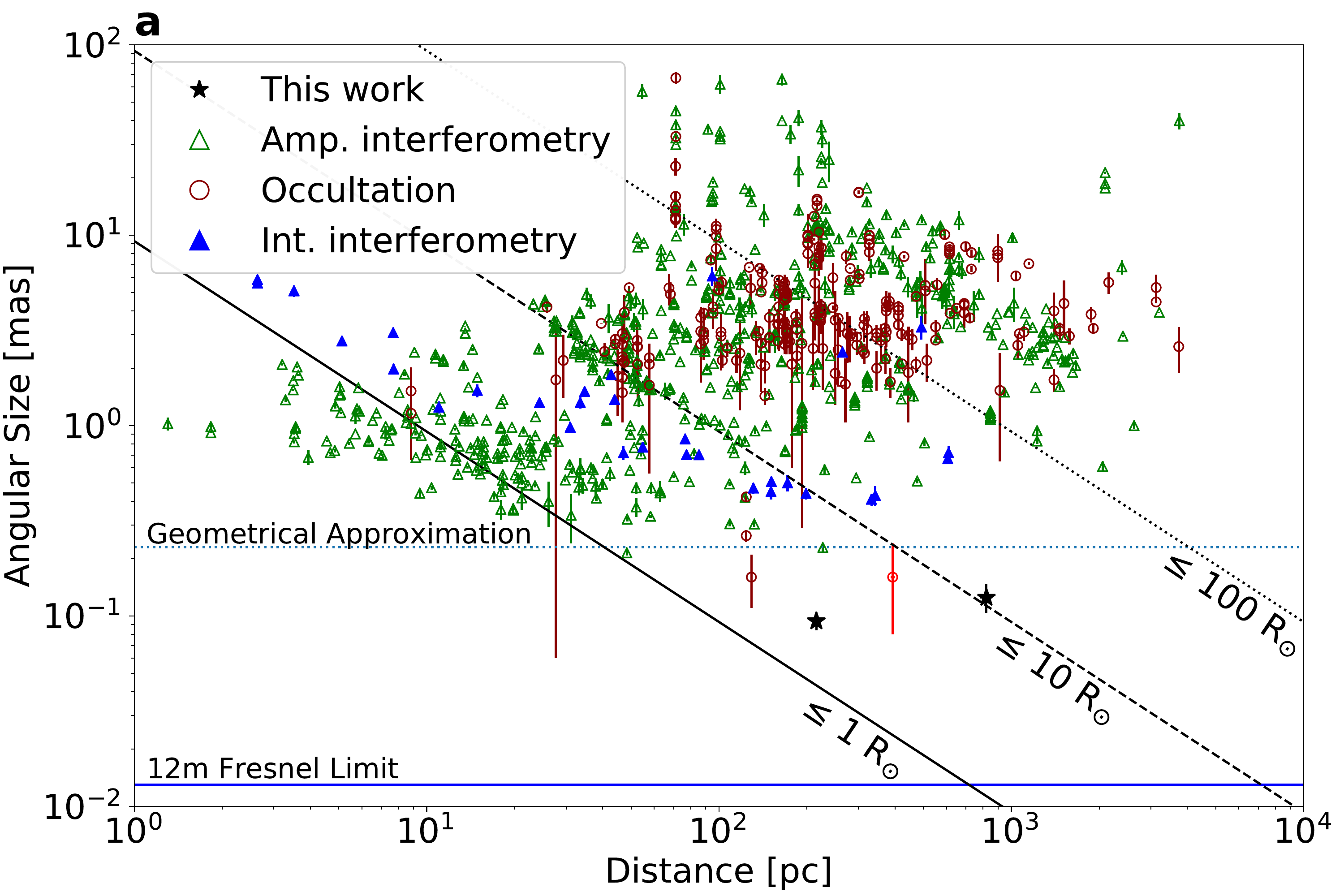} 
\includegraphics[width=0.66\linewidth]{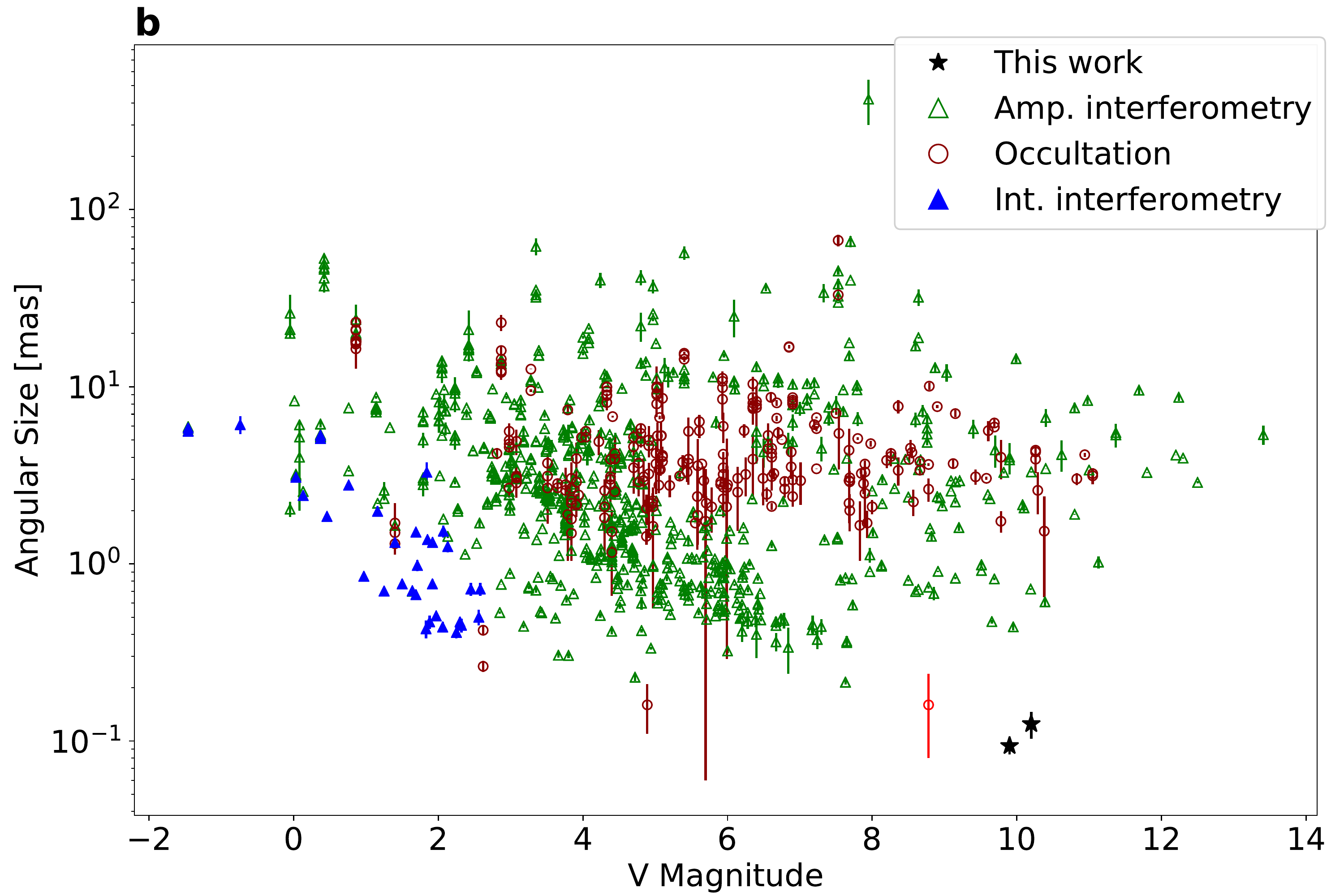} 
\end{center}
\caption{\textbf{Comparison to the available directly measured stellar angular size measurements.} 
\textbf{a}: The angular size as a function of distance for all stars with direct measurements\cite{JMDC}. 
The asteroid occultation measurements are marked with stars, this work in black, the (3)\,Juno occultation in red; amplitude interferometry measurements by open green triangles; intensity interferometry by blue triangles; and all other occultation measurements by open dark red circles. 
The solid, dashed and dotted lines respectively show the expected value for a 1, 10 and 100 solar radius star. 
The blue solid line gives the theoretical limit for discriminating between a point-like source and a resolved star by its Fresnel diffraction and the blue dashed line the region where the diffraction pattern completely disappears for a geometrically resolved star. 
\textbf{b}: As before, but for angular size as a function of apparent magnitude. 
Errors for this work are the 68\% confidence level, all others are taken from the respective catalogue entry\cite{JMDC}.
}
\label{fig:SizeVsDist}
\end{figure}

The angular size as a function of distance for all stars with direct size measurements to date is shown in Figure~\ref{fig:SizeVsDist}\textbf{a}. 
The measurements presented here represent a factor of 10 improvement in angular size resolution compared to the standard lunar occultation method and are also a factor of at least two smaller than available interferometric size measurements\cite{JMDC, Chelli16}. 
Remarkably, this places these direct measurements of the angular size in the same region of parameter space as the empirically derived estimates of angular size for stars that are being used as unresolved point sources by interferometers for calibration\cite{JMDC, Chelli16}. The closest measurements in angular scale, again larger by a factor of two, come from the occultation of the +2.5 m$_V$ $\beta$\,Sco system by Jupiter\cite{betaSco}, and the occultation of the star SAO\,115946 by the asteroid (3)\,Juno\cite{Juno} in the 1970s. 
Both of those measurements benefited similarly from the large distance to the occulting object, but were extremely limited by high levels of scintillation noise in the telescope's data leading to large ($\simeq50\%$) uncertainties (and also limited to an observation of a very bright, rare, object in the case of $\beta$\,Sco). 
Our uncertainty ($\sim$10\% level) is currently limited by the signal to noise ratio within the diffraction fringes from our initial relatively simple broadband set up. 
Implementing a narrower band filter would be a way to reduce dilution of the fringes and potentially further improve the accuracy of these measurements.

Asteroid occultation shadows regularly pass over the Earth's surface, with the potential number of occulted stars per year exponentially increasing with the apparent magnitude as you progress to fainter stars.
The shadow paths are predicted by combining star catalogues with the orbital ephemerides of known asteroids with a precision usually comparable to the asteroid size. 
The problem with exploiting these occultations is that only $\simeq7\%$ of them have a $\geq20\%$ chance of actually being observed from any fixed location, making them difficult to catch with the kinds of large, non-portable, telescope that are necessary to resolve the fast moving deviations in the shadow signal over the scintillation noise introduced by the Earth's atmosphere. 
However, a telescope capable of detecting an occultation of a 10th magnitude star can view, on average, 5 viable occultations per year, increasing to almost 1 per week for occultations of 13th magnitude stars. 
The faintness of the objects that we have observed, shown in Figure~\ref{fig:SizeVsDist}\textbf{b}, also represents nearly an order of magnitude increase in distance, when compared to stars of similar radii, that have had their angular size directly measured. 
This means we are not limited to nearby, bright objects and so greatly increases the volume of space, and variety of stars, that can be sampled through this technique. 
In summary, the improved sensitivity provided by IACTs greatly increases the chances of observing an asteroid occultation from a fixed site to a rate sufficient to obtain a viable sample for population studies for use in areas such as stellar evolution modelling\cite{Mozurkewich2003} or transiting exoplanet radius measurements\cite{vonBraun2014}. 
The imminent construction of the Cherenkov Telescope Array\cite{CTA,CTAWP} opens the way for many exciting opportunities in high time resolution precision photometry with IACT arrays in the near future.

\begin{addendum}
 
 \item[Correspondence] Correspondence and requests for materials should be addressed to T.~Hassan~(email: tarek.hassan@desy.de) or M.~Daniel (email:michael.daniel@cfa.harvard.edu).
 \item[Acknowledgements] This research is supported by grants from the U.S. Department of Energy Office of Science, the U.S. National Science Foundation and the Smithsonian Institution, by NSERC in Canada, and by the Young Investigators Program of the Helmholtz Association. We acknowledge the excellent work of the technical support staff at the Fred Lawrence Whipple Observatory and at the collaborating institutions in the construction and operation of the instrument. 
 This work has made use of data and updates by S. Preston from \url{http://www.asteroidoccultation.com}; 
 data from the JPL Small-Body Database browser at \url{http://ssd.jpl.nasa.gov}; 
 data from the European Space Agency (ESA) mission {\it Gaia} (\url{https://www.cosmos.esa.int/gaia}), processed by the {\it Gaia} Data Processing and Analysis Consortium (DPAC,  \url{https://www.cosmos.esa.int/web/gaia/dpac/consortium}); 
 and the SIMBAD database, operated at CDS, Strasbourg, France.
 The authors want to acknowledge discussions with S. Cikota and J. Cortina.
 \item[Author Contributions] All authors contributed equally to the operation of the VERITAS telescopes. M.D. conceived the enhanced current monitor system used in these observations; T.H. proposed the occultation observations; A.J., D.W., T.W., J.Q., A.B. took the observations; M.D., T.H. and N.M. reduced and analysed the data; M.D. and T.H. wrote the main paper and methods section.
 \item[Competing Interests] The authors declare that they have no
competing financial interests.
\end{addendum}

\clearpage

\section*{Methods}
\section*{Instrument, observations \& data reduction}
The Very Energetic Radiation Imaging Telescope Array System (VERITAS) is sited at the FLWO, with its primary research focus in the area of particle astrophysics relating to the ground-based detection of very-high-energy (VHE) $\gamma$-rays\cite{Weekes96,2009ARAA}. 
The system is comprised of four 12\,m diameter segmented reflectors each viewed by a camera of 499 photomultiplier tubes with a 0.15$^\circ$ pixel field of view closely matching the optical point spread function\cite{VERITAS}. 
A set of up to 16 pixels per camera have been recently upgraded to monitor the DC light level in the field of view with a commercial DATAQ DI-710-ELS DC voltage datalogger with 14-bit resolution and sampling rates up to 4,800\,Hz. 
For the Imprinetta proof-of-principle observation we limited the data throughput to 300\,Hz (3\,ms between samples) and for the Penelope observation this was raised to 2,400\,Hz (0.4\,ms between samples). 
The datalogger and the Cherenkov data acquisition can be used for
simultaneous optical and gamma-ray coverage. 
The datalogger only has a coarse non-synchronised clock for timestamping the samples, so background pixels were used to compare the time of shooting star events moving through the Cherenkov camera, which has its events timestamped with a GPS clock. 
This allowed absolute timing corrections accurate to the level of 0.02\,s to be made to the DC light level samples. 
For a telescope of diameter $D$ and sampling time $t$, the intensity fluctuations from scintillation noise\cite{Scintillation} scale as $\Delta{I}/I \propto D^{-2/3}/\sqrt{t}$, which means the 12\,m VERITAS telescopes with millisecond sampling should have noise levels at least 20 times lower than a portable 50\,cm telescope equipped with high frame rate video ($\sim$60\,Hz). 

\section*{Diffraction pattern analysis}
An asteroid intersecting the line of sight between the observer and a star casts a shadow moving at its projected velocity. 
The edge of this shadow, instead of having a sharp boundary, shows a diffraction pattern produced by the asteroid limb. 
Taking into account the distance ($\sim$ 4 $\times$ 10$^{11}$ m) and size of these asteroids (tens of km), the diffraction pattern produced, as a first approximation, is equivalent to that of an infinite straight edge\cite{disk_vs_straight_edge}, which in the case of a point source can be expressed as 
\begin{equation}
\label{eq:diffraction}
I(x) = I_{BG} + I_0 \left( \left| \int_{-\infty}^{x} \cos (\frac{\pi r^2}{L \lambda}) dr \right|^2 + \left| \int_{-\infty}^{x} \sin (\frac{\pi r^2}{L \lambda}) dr \right|^2 \right),
\end{equation}
where $r$ and $x$ are the distance to the centre and edge of the geometric shadow respectively, $I_0$ is the signal intensity produced by the star, $L$ the distance to the asteroid, and $\lambda$ the wavelength of the photons collected. 
This shadow is detected by VERITAS as it passes over FLWO with speed $v$ (11.8 and 2.2 km/s for Imprinetta and Penelope on their respective dates), measuring the diffraction pattern imprint as a function of time. 
A star of diameter $d$ at a distance $D$ has an angular diameter $\delta = 2 \arctan(\frac{d}{2D})$. 
This size is projected into the asteroid shadow to a size $d_{proj} = L\tan(\delta)$, producing a smearing of the pattern on such a scale. 
Therefore if $d_{proj}$ is comparable to the Fresnel scale $\sqrt{L \lambda}/2$, the distinct smearing of the pattern (in the simplest case, assuming a uniform disc) allows us to directly measure the star diameter. 
The power of asteroid occultation over lunar occultation relies on the difference in distance between the Moon and main belt asteroids (a factor $\sim$10$^3$), which translates into a ratio between the projected star size and the Fresnel scale $\sim30 \times$ larger. 

Apart from the size of the star, several effects influence the theoretical fringe pattern expected from these occultations\cite{measurement_lunar_occults}: the optical bandwidth detected smears the pattern (mainly after the first dip), while the asteroid velocity $v$ and the occultation angle $\theta_{occult}$ (angle between the asteroid trajectory and occulting surface) modify the time scale of the detected pattern by $v\cos(\theta_{occult})$. 
Given the negligible uncertainty of main belt asteroid trajectories, the principal uncertainty on the theoretical diffraction pattern considered in the analysis is the optical bandwidth of the VERITAS detectors. 
The main parameters affecting the measured photons are the star spectrum, the atmospheric transmission\cite{Atmosphere}, the mirror reflectivity\cite{Mirror} and the photomultiplier tube's quantum efficiency\cite{Gazda2016} (all wavelength dependent). 
As IACTs rely heavily on detailed Monte Carlo simulations of both the atmosphere and ray tracing, these parameters are well understood. 
Spectral templates\cite{pickles1998, Kesseli2017} were used to model both occulted stars, corresponding to the spectral types K3 and G0 for Imprinetta and Penelope occultation respectively. 
Combining all these contributions, the resulting optical passband is a 120\,(140)\,nm  band centred at 470\,(450)\,nm for Imprinetta\,(Penelope). This effect was included in the analysis by convolving the weighted monochromatic diffraction patterns over the resulting optical bandpass.
The systematic uncertainty of this distribution was tested, modifying these contributions within their expected variance. 
The first fringe (the most constraining part of the pattern) is barely affected (below $\sim$1\% in flux) while starting from the second one uncertainties reach up to 5\% in flux.

A $\chi^2$ minimisation method was applied to find the model best describing the observed patterns. 
Each pattern was fitted independently (i.e., four ingress and four egress for Imprinetta) leaving the time of occultation and the occultation angle as free parameters. 
Only the region of time in which the theoretical point-source diffraction fringes are expected to deviate from the uniform flux profile was used to calculate the $\chi^2$ values. 
A full parameter profiling over the free parameters was performed in fixed steps of assumed constant star sizes (68 steps between 0.01 and 0.35 mas) for each pattern. All distributions from the same occultation were then combined and final p-values were calculated from the resulting $\chi^2$. 
In the case of the Penelope occultation, only data from two telescopes were used, as the other two were being used to test alternative observing strategies, which resulted in data incompatible for use in this analysis.

\begin{addendum}
 \item[Data Availability] The datasets generated and/or analysed in this study are available from the corresponding authors on request.
 \item[Code Availability] The computer code used to analyse the data in this study is available from the corresponding authors on request. 
\end{addendum}

\end{document}